\documentclass[aps,prd,preprint]{revtex4-2}

\usepackage[T1]{fontenc}          
\usepackage[latin9]{inputenc}     

\usepackage{babel}
\AtBeginDocument{}

\usepackage{amsmath}
\usepackage{amssymb}

\usepackage{bm}
\usepackage{hyperref}
\usepackage{graphicx} 
\makeatletter
\newcommand{\lyxmathsym}[1]{%
  \ifmmode
    \begingroup
      \def\b@ld{bold}%
      \text{\ifx\math@version\b@ld\bfseries\fi#1}%
    \endgroup
  \else
    #1%
  \fi
}
\makeatother

\begin{document}
\title{Covariant Quantum Measurements and Stochastic Dynamics on Representation Space}

\author{Naeem Shahid}
\affiliation{Department of Physics, Pennsylvania State University, New Kensington,
	PA 15068, USA.}
\begin{abstract}
We develop a framework for group--covariant quantum measurements in which measurement--induced transitions between irreducible representation sectors are described by a stochastic process on representation space. Starting from the Peter--Weyl decomposition, we construct covariant measurement operators from irreducible tensor operators and show that, for symmetry--invariant states, the measurement channel reduces to a Markov process on the representation graph. We further show that analyticity of the measurement operator constrains the detector spectrum through Sugiura's theorem, motivating a class of exponentially decaying detector models. Specializing to SU(2), we obtain the transition kernel in closed form, establish reversibility and the associated invariant measure, and derive a continuum Fokker--Planck description of the induced dynamics. Analytical predictions for the drift and diffusion coefficients are found to agree with numerical simulations. Our results provide a stochastic description of repeated covariant quantum measurements on representation space.
\end{abstract}
\maketitle

\section{Introduction}

Group covariant quantum measurements are a class of measurements that
respects a symmetry group. When a quantum state is transformed by
a group action, the measurement outcomes are also transformed such
that the physically equivalent outcomes cannot be distinguished by
the experiment.

Mathematically, a quantum measurement is described by a positive operator--valued
measure (POVM) $M(x)$. If $G$ is a symmetry group with a unitary
representation $U(g)$ acting on the system's Hilbert space, then the measurement
is covariant if,

\begin{equation}
	U(g)M(X)U(g)^{\dagger}=M(gX),\quad\forall\,g\,\in G.
\end{equation}

Here, $gX$ denotes the transformed outcomes set under the action
of the group.

Covariant measurements play a central role in quantum estimation theory \cite{Holevo_1979,Barnett_1989,Ariano2003,Paris:2009aa,Holevo_2011}, where the goal is to estimate an unknown parameter encoded through a symmetry transformation. Examples include optimal estimation strategies, quantum Fisher information, and phase or displacement estimation. For many estimation problems with symmetric priors and cost functions, covariant POVMs are provably optimal, allowing the search for an optimal measurement to be restricted to the class of covariant measurements without loss of performance.

In this work we consider covariant measurements from a different perspective. Rather than focusing on parameter estimation, we study how a covariant measurement redistributes probability among the irreducible representation sectors of the underlying symmetry group. Working within the Peter--Weyl decomposition, we show that every covariant measurement naturally induces a stochastic process on the representation graph, with transition probabilities determined by the intertwining structure of the measurement operators. Furthermore, analyticity of the measurement operator imposes exponential constraints on the detector coefficients through Sugiura's theorem.

The construction is formulated for arbitrary compact Lie groups and does not depend on the explicit form of the intertwiners. However, the resulting transition kernels depend on the representation theory of the underlying group. To obtain explicit analytical expressions we specialize to the group SU(2), where the intertwiners are given by Clebsch--Gordan coefficients. For a physically motivated family of detector models we derive the corresponding transfer matrix, establish reversibility and the associated invariant measure, and obtain a continuum Fokker--Planck description of the induced dynamics on representation space.

The present work is therefore not intended as a model of a particular laboratory measurement. Instead, it develops a mathematical framework for symmetry--covariant quantum instruments whose repeated action generates transport and diffusion on representation space. Although the explicit calculations are carried out for SU(2), the general formalism extends to arbitrary compact groups, with the corresponding representation theory determining the detailed form of the stochastic dynamics.

\section{Covariant measurement framework}

\subsection{Intertwiners }

Let G be a compact Lie group acting unitarily on a Hilbert space $\mathcal{H}$.
The unitary representation is denoted by,

\begin{equation}
	U:G\to\mathcal{B}(\mathcal{H}),\quad g\mapsto U(g).
\end{equation}

For compact $G$, the Peter-Weyl theorem guarantees that $\mathcal{H}$
decomposes as,

\begin{equation}
	\mathcal{H}=\bigoplus_{\pi \in \hat{G}}\mathcal{H}_{\pi}\otimes\mathbb{C}^{m_{\pi}},
\end{equation}

where $\hat{G}$ is the set of irreducible representations (irreps)
of $G$, and each $\mathcal{\ensuremath{H}}_{\pi}$ carries an irrep
of dimension $d_{\pi}$. $m_{\pi}$ denotes the multiplicity of the irrep $\pi$.

The operator algebra $\mathrm{End}(\mathcal{H})$ carries the adjoint
action,

\begin{equation}
	X\mapsto U(g)XU(g)^{\dagger},\quad g\in G.
\end{equation}

Under this action, it decomposes into irreducible invariant subspaces,
\begin{equation}
	End(\mathcal{H})\cong\bigoplus_{\pi\in\widehat{G}}V_{\pi}\otimes\mathbb{C}^{\tilde{m}_{\pi}},
\end{equation}

where $V_\pi$ is a $d_\pi$--dimensional carrier space for a copy of the irrep $\pi$, and $\tilde{m}_\pi$ is its multiplicity in $\mathrm{End}(\mathcal{H})$.

For each irrep ($\pi\in\hat{G}$), let,

\begin{equation}
	T_{m,\lambda}^{(\pi)},\quad m=1,\ldots,d_{\pi}\label{eq:intertwiner_gen},
\end{equation}

denote an irreducible tensor--operator basis transforming according
to $\pi$,

\begin{equation}
	U(g)T_{m,\lambda}^{(\pi)}U(g)^{\dagger}=\sum_{r=1}^{d_{\pi}}D_{rm}^{(\pi)}(g)T_{r,\lambda}^{(\pi)},
\end{equation}

where $D^{(\pi)}(g)$ is the matrix of the irrep $\pi$. The label,
\begin{equation}
	\lambda\equiv(\alpha,\beta,\mu),\quad\mu=1,\cdots N_{\alpha\pi}^{\beta},
\end{equation}
specifies the multiplicity of the intertwining map, where,
\begin{equation}
	N_{\alpha\pi}^{\beta}=\dim\,Hom_{G}(\mathcal{H}_{\alpha}\otimes\mathcal{H}_{\pi},\mathcal{H}_{\beta}),
\end{equation}

is the multiplicity of the irrep $\mathcal{H}_{\beta}$ in the decomposition of $\mathcal{H}_{\alpha}\otimes\mathcal{H}_{\pi}$.

The tensor operators form a basis of intertwining maps between irreducible
representation sectors. Their explicit realization depends on the
group. For example, for U(1) the intertwining coefficients are trivial,
whereas for SU(2) they are given by the Clebsch--Gordan coefficients.
Throughout this work the general formalism is developed independently
of the explicit realization, while Sections III onward specialize
to SU(2).

\subsection{Measurement operators}

A family of measurement operators $\{M_{g}\}_{g\in G}$ is $G$--covariant
if,

\begin{equation}
	U(h)M_{g}U(h)^{\dagger}=M_{hg},\qquad\forall\,h,g\in G.
\end{equation}

By the Peter--Weyl theorem, any square--integrable function on $G$
decomposes in the matrix elements $D_{mn}^{\pi}(g)$ of the irreps.
The measurement operator can then be written as,

\begin{eqnarray}
	M_{g} & = & \sum_{\pi}\sum_{m,n}D_{mn}^{(\pi)}(g)A_{nm}^{(\pi)},\\
	A_{nm}^{(\pi)} & = & \sum_{\lambda}c_{n\lambda}^{(\pi)}T_{m,\lambda}^{(\pi)}.
	\label{eq:M_op}
\end{eqnarray}

The corresponding POVM elements $M_{g}^{\dagger}M_{g}$ satisfy,

\begin{equation}
	\int_{G}d\mu(g)M_{g}^{\dagger}M_{g}={\bf I},\label{eq:comp_of_M}
\end{equation}
where $d\mu(g)$ is the Haar measure on $G$.

The coefficients $c_{n\lambda}^{(\pi)}$ encode the response of the
detector. They specify how strongly each tensor channel contributes
to the measurement and therefore determine the transition probabilities
between irreducible sectors.

The updated state of the system can then be computed as,
\begin{equation}
	\bar{\rho}=\int_{G}d\mu(g)M_{g}\rho M_{g}^{\dagger}.
\end{equation}

Finally, if $\Pi_{\alpha}$ denotes the projector onto the irreducible subspace $\mathcal{H}_{\alpha}$,
the occupation probability of the irrep $\alpha$ after the measurement
is,

\begin{equation}
	p_{\alpha}=\mathrm{Tr}(\Pi_{\alpha}\bar{\rho}).
\end{equation}

Our objective here is to characterize how a covariant measurement redistributes
probability among irreducible representation sectors. This general construction will now be specialized to SU(2), where explicit tensor operators and transition
kernels can be obtained in closed form.

\section{SU(2) Specialization}

The general construction in the last section becomes completely explicit for SU(2), where
the intertwining operators in Eq. (\ref{eq:intertwiner_gen}) are determined
by Clebsch--Gordan coefficients.

\begin{equation}
	T_{M;jj'}^{(J)}=\sum_{m,m'}C_{jm;JM}^{j'm'}|j'm'\rangle\langle jm|.
\end{equation}

These intertwiners satisfy the Hilbert--Schmidt orthogonality relation
(Appendix \ref{app:hs_ortho}).

\begin{equation}
	\mathrm{Tr}\left(T_{Q;kk'}^{(K)}T_{M;jj'}^{(J)\dagger}\right)=\frac{d_{j^{\prime}}}{d_{J}}\delta_{jk}\delta_{j^{\prime}k^{\prime}}\delta_{JK}\delta_{QM}.\label{eq:HS_ortho}
\end{equation}

Note that we are using the standard notation by identifying the incoming
irrep $\alpha$ with $j$ while outgoing irrep $\beta$ with $j^{\prime}$.
For SU(2), the Clebsch--Gordan decomposition is multiplicity free,
i.e. $N_{\alpha\pi}^{\beta}\in\{0,1\}$, so the multiplicity label
$\mu$ may be omitted. 

Similarly, the measurement operator (Eq. (\ref{eq:M_op})) in the standard SU(2) notations
is,
\begin{eqnarray}
	M_{g} & = & \sum_{J}\sum_{M,N}D_{MN}^{(J)}(g)A_{NM}^{(J)},\\
	A_{NM}^{(J)} & = & \sum_{j,j'}c_{N;jj'}^{(J)}T_{M;jj'}^{(J)}.\label{eq:PW_exapnd}
\end{eqnarray}

In order to guarantee that the measurement describes all possible
outcomes and that the total probability is always 1, the POVM normalization
condition (Eq. (\ref{eq:comp_of_M})) must hold.

This imposes (Appendix \ref{app:M_complete}) a constraint on the detector
coefficients $c_{N;jj'}^{(J)}$, which will be specified in the next section.

\begin{equation}
	\sum_{J,j',N}\frac{d_{j^{\prime}}}{d_{J}d_{j}}|c_{N;jj'}^{(J)}|^{2}=1,\quad\forall\:j.\label{eq:comp_cond_M}
\end{equation}

Here each $d_{j},d_{j^{\prime}}$ and $d_{J}$ are the dimension of
input, output and the detector space respectively. 

For a general density operator the covariant measurement induces a
quantum channel on the full operator algebra. Restricting to Casimir--diagonal
states closes the dynamics on the irrep populations, allowing the
evolution to be described by a classical Markov process.

Consider an initial state that is block diagonal in the irreducible
decomposition,
\begin{equation}
	\rho=\sum_{q}\frac{p_{q}\Pi_{q}}{d_{q}},\quad d_{q}=2q+1,
\end{equation}
then the probability of being in irrep $r$ after measurement ($\mathrm{Tr}(\Pi_{r}\bar{\rho})$)
can be computed as,
\begin{equation}
	p_{r}=\sum_{q}\frac{p_{q}d_{r}}{d_{q}}\sum_{J,N}\frac{1}{d_{J}}|c_{N;qr}^{(J)}|^{2}=\sum_{q}p_{q}K_{rq}.
\end{equation}

This expression immediately leads to the transfer matrix,

\begin{equation}
	K_{rq}=\sum_{J,N}\frac{d_{r}}{d_{q}d_{J}}|c_{N;qr}^{(J)}|^{2}\label{eq:transfer_K}
\end{equation}

such that,
\begin{equation}
	\sum_{r}K_{rq}=\sum_{r,J,N}\frac{d_{r}}{d_{q}d_{J}}|c_{N;qr}^{(J)}|^{2}=1.\label{eq:K_col_stochastic}
\end{equation}

Where we used the completeness condition obtained above, Eq. (\ref{eq:comp_cond_M}). 

It shows that the transfer matrix is column stochastic. Moreover, the transfer
matrix defines an effective geometry on representation space with
vertices given by irreps $j=0,1/2,1,3/2,\cdots$ and edges fixed by
allowed values of the intertwiners. The measurement itself induces
diffusion with weights fixed by detector coefficients, and the dynamics
set by covariance.

The detector coefficients are the only free ingredients of this formalism.
Different choices correspond to different physical measurement devices.
The remainder of the paper investigates one analytically motivated
detector whose transition probability decays exponentially with transferred
tensor rank.

\section{Analytic constraints on detector coefficients}

If the measurement operator $M_{g}$ is an analytic operator--valued
function on SU(2), then its Peter--Weyl coefficients satisfy Sugiura's
exponential estimate. Since these coefficients are linear combinations
of the tensor operators weighted by $c_{N;jj'}^{(J)}$, analyticity
imposes exponential decay on the coefficient arrays \cite{Sugiura1971,Gangolli_1988}. 

The theorem states, given
the Peter--Weyl expansion Eq. \ref{eq:PW_exapnd}, a function $M$
is real analytic on $G$ if and only if there exist constants $C,t>0$
such that,

\begin{equation}
	\left\Vert A^{(J)}\right\Vert \le Ce^{-t\left|\lambda_{J}\right|},\quad\forall\,J.
\end{equation}

Where $\lambda_{J}$ is the highest weight of the irreducible representation
$J$.

Using Hilbert--Schmidt orthogonality relation Eq. \ref{eq:HS_ortho},
we get,

\begin{equation}
	\left\Vert A^{(J)}\right\Vert ^{2}=\mathrm{Tr}\left(\mathcal{A}_{NM}^{(J)\dagger}\mathcal{A}_{NM}^{(J)}\right)=\sum_{j,j'}\frac{d_{j^{\prime}}}{d_{J}}\left|c_{N;jj'}^{(J)}\right|^{2}.
\end{equation}

Now the Sugiura's theorem yields,

\begin{equation}
	\left[\sum_{j,j'}\frac{d_{j^{\prime}}}{d_{J}}\left|c_{N;jj'}^{(J)}\right|^{2}\right]^{1/2}\le Ce^{-t|\lambda_{J}|},
\end{equation}
where for SU(2) $|\lambda_{J}|=2J$.

Sugiura's theorem constrains only the asymptotic decay of the collective Fourier coefficients, but does not uniquely determine the individual transition amplitudes $c_{N;qr}^{(J)}$. The microscopic distribution of the Fourier weight among different irreducible--sector transitions therefore remains model dependent. We consequently introduce the following phenomenological detector model,
\begin{equation}\label{detector_ansatz}
	 \sum_N|c_{N;qr}^{(J)}|^2 = \chi_q d_J e^{-\beta J},
\end{equation}

which assigns an exponentially decreasing probability to higher--rank tensor components while remaining consistent with the asymptotic exponential decay required by Sugiura's theorem.

Here $\beta$ controls the detector bandwidth. This ansatz is consistent
with Sugiura's theorem whenever $\beta>4t$. If the detector suppresses tensor ranks faster than the minimum rate
dictated by the analyticity radius $4t$, then the detector defines
an admissible analytic covariant measurement.

\section{Reversibility and invariant measure}

Substituting the detector ansatz Eq. (\ref{detector_ansatz}) into
the transfer matrix Eq. (\ref{eq:transfer_K} ), the transition kernel
becomes,

\begin{equation}
	K_{rq}=\frac{1}{Z_{q}(\beta)}\left[d_{r}e^{-\beta|r-q|}\left\{ 1-e^{-\beta(r+q-|r-q|+1)}\right\} \right].
\end{equation}

The normalization factor $Z_{q}(\beta)$ can be found using the column--stochastic condition from Eq. \ref{eq:K_col_stochastic},

\begin{equation}
	Z_{q}(\beta)=d_q\frac{1+e^{-\beta/2}}{1-e^{-\beta/2}}.
\end{equation}

\begin{figure*}[t]
	\includegraphics[width=1\columnwidth]{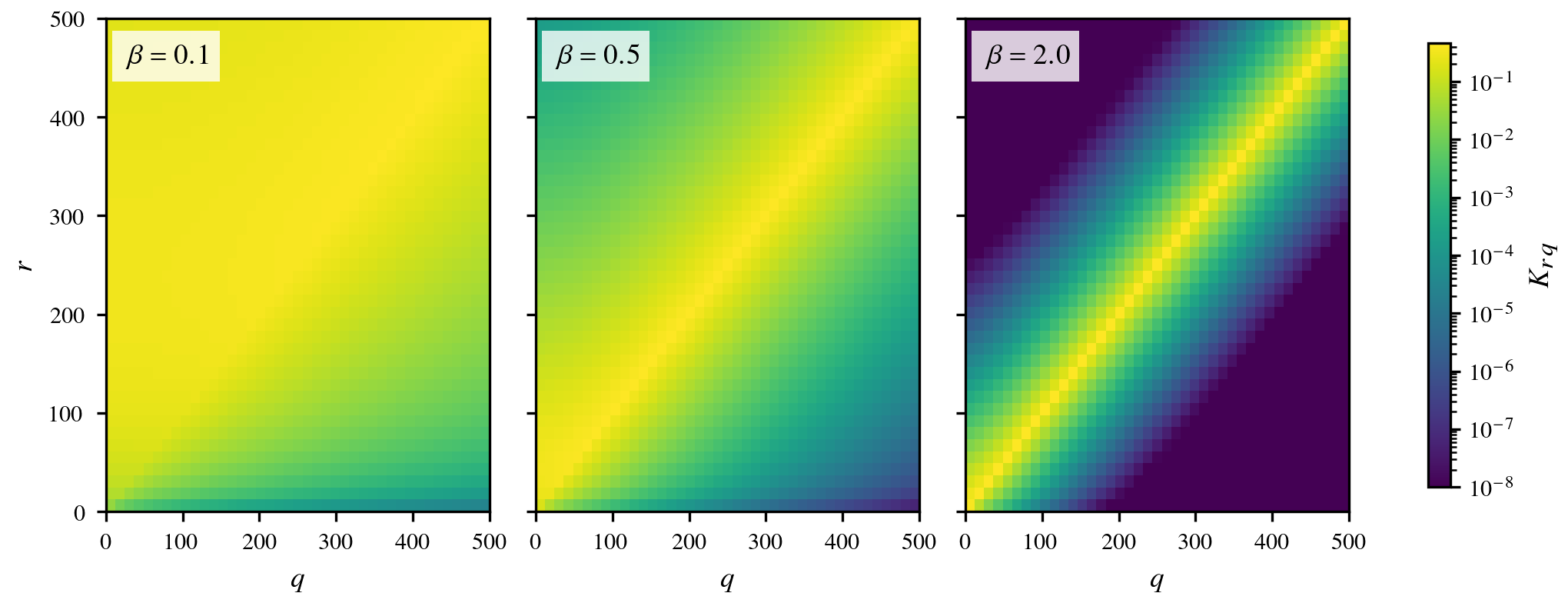}
	\caption{Heat maps for the transition kernel for several values of the detector bandwidth parameter $\beta$ .}
	\label{fig:heat_map}
\end{figure*}
\begin{figure*}[t]
	\includegraphics[width=1\columnwidth]{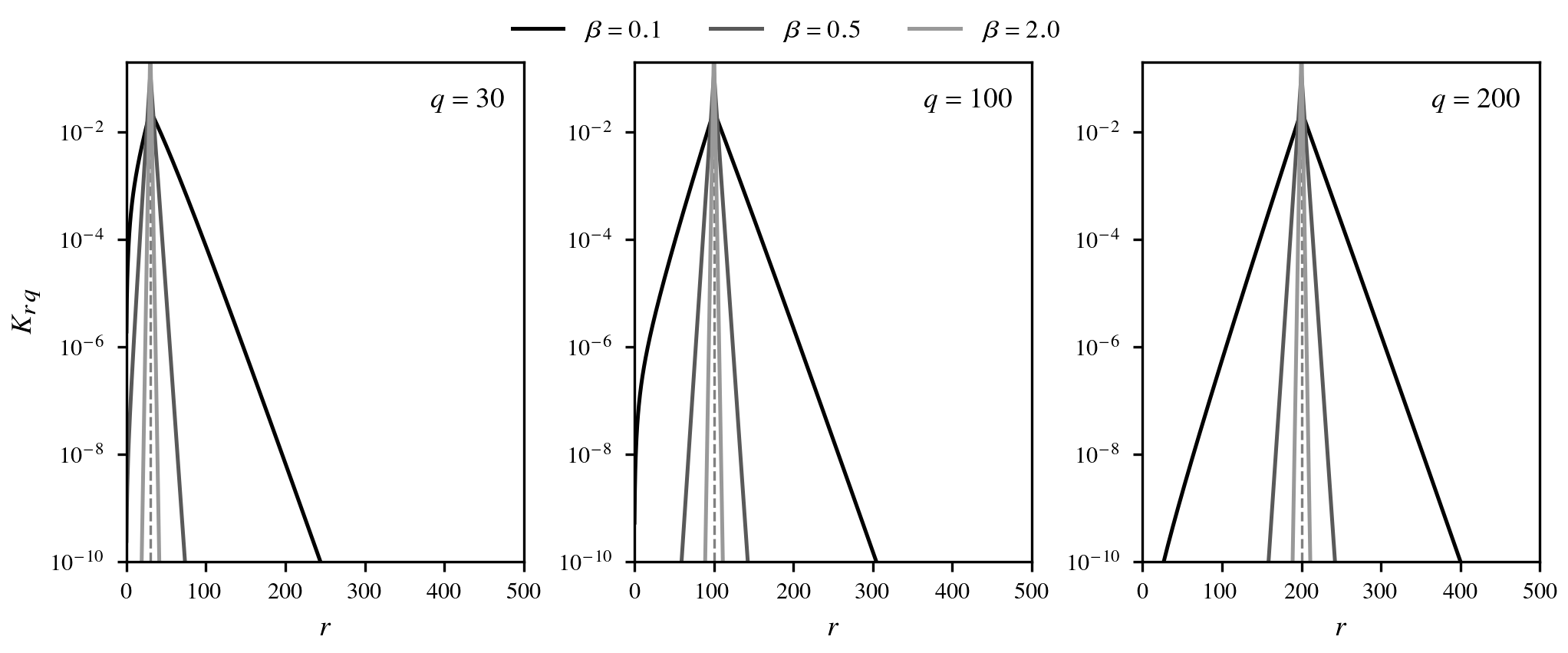}
	\caption{One--dimensional slices of the transition kernel for fixed initial representations for several values of the detector bandwidth parameter $\beta$ .}
	\label{fig:K_vs_r}
\end{figure*}

Fig. \ref{fig:heat_map} illustrates the transition kernel for several values
of the detector bandwidth parameter $\beta$. For small $\beta,$
the kernel is broad, allowing transitions between widely separated
irreducible representations. As $\beta$ increases, the transition
probability becomes increasingly localized around the diagonal $r=q$,
indicating that long--range transitions are exponentially suppressed.
The detector therefore interpolates continuously between a highly
nonlocal measurement ($\beta\ll1$) and an almost local measurement
($\beta\gg1$). The corresponding heat maps show that increasing $\beta$
continuously narrows the interaction band on the representation graph.

Fig. \ref{fig:K_vs_r} shows the one--dimensional slices of the transition kernel as a function of the final representation $r$ for several fixed initial representations $q$. Increasing $\beta$ progressively suppresses long--range transitions, producing an increasingly localized kernel centered at $r=q$. The nearly identical profiles for $q= $ 30,100, and 200 indicate that boundary effects become negligible for large representations.

The covariant measurement induces a reversible Markov chain on the
SU(2) representation graph. It admits the invariant measure $\mu_{r}\propto d_{r}^{2}$,
which satisfies the detailed balance relation $\mu_{q}K_{rq}=\mu_{r}K_{qr}$.
Since $\sum_{r}\mu_{r}=\sum_{r}(2r+1)^{2}=\infty$, the invariant
measure has infinite total mass and therefore cannot be normalized
into a probability distribution. Consequently, the chain possesses
no normalizable stationary probability distribution on the infinite
representation graph. A stationary state exists only after introducing
either a finite representation cutoff or an additional confining mechanism.

\section{Continuum limit and Fokker--Planck description}

\begin{figure*}[t]
	\includegraphics[width=1\columnwidth]{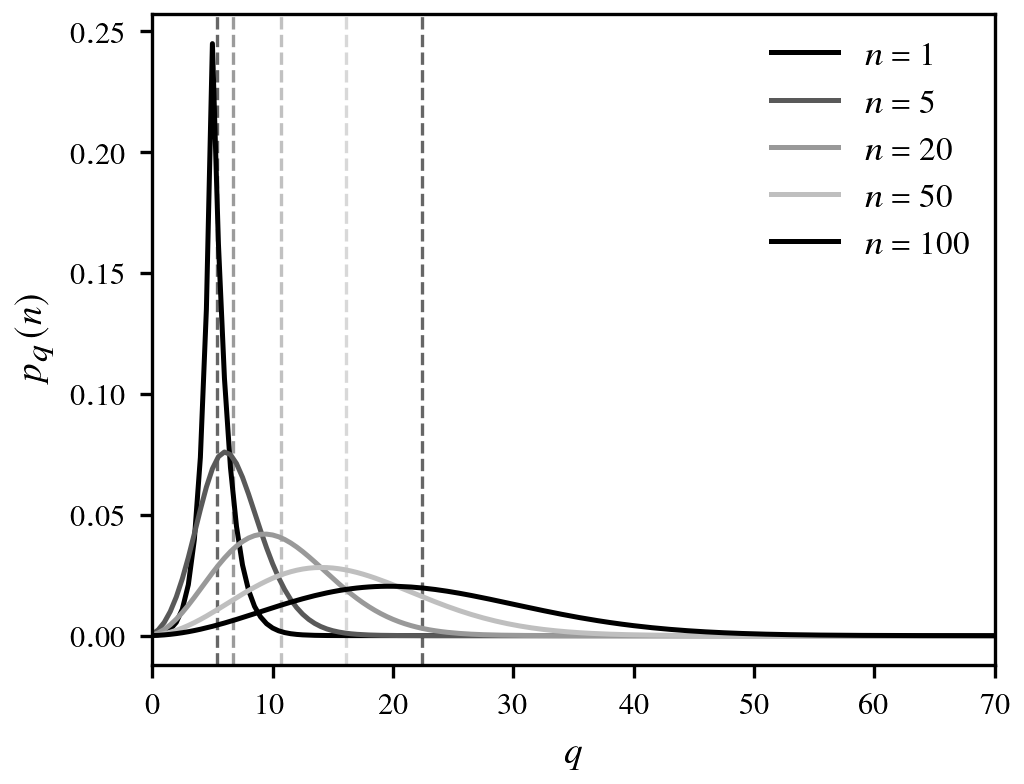}
	\caption{The evolution of the probability distribution under repeated applications
		of the measurement channel for $\beta=1$}
	\label{fig:prob_n}
\end{figure*}

The evolution of the probability distribution under repeated applications
of the measurement channel is shown in Fig. \ref{fig:prob_n} for an intermediate
value of $\beta=1$. As the measurement is applied repeatedly, the
initially localized distribution broadens while its maximum gradually
shifts toward larger irreducible representations. This reflects the
combined action of diffusion in representation space together with
the weak outward drift induced by the detector.

We can compute the drift and diffusion induced by the measurement,
\begin{eqnarray*}
	A(q) & = & \sum_{r}(r-q)K_{rq},\\
	B(q) & = & \sum_{r}(r-q)^{2}K_{rq},
\end{eqnarray*}

which are respectively the first and second jump moments of the transfer
kernel.

In the continuum limit, the Markov chain is described by the Kramers--Moyal
expansion. Truncating the expansion after the second jump moment yields
the Fokker--Planck equation,
\begin{equation}
	\frac{\partial P(q,n)}{\partial n}=-\frac{\partial}{\partial q}\!\left[A(q)P(q,n)\right]+\frac{1}{2}\frac{\partial^{2}}{\partial q^{2}}\!\left[B(q)P(q,n)\right].
\end{equation}

The corresponding drift and diffusion coefficients are,

\begin{eqnarray*}
	A(q) & = & \frac{e^{-\beta/2}}{2q(1-e^{-\beta/2})^{2}},\\
	B(q) & = & \frac{e^{-\beta/2}}{2(1-e^{-\beta/2})^{2}}.
\end{eqnarray*}

\begin{figure*}[t]
	\includegraphics[width=1\textwidth]{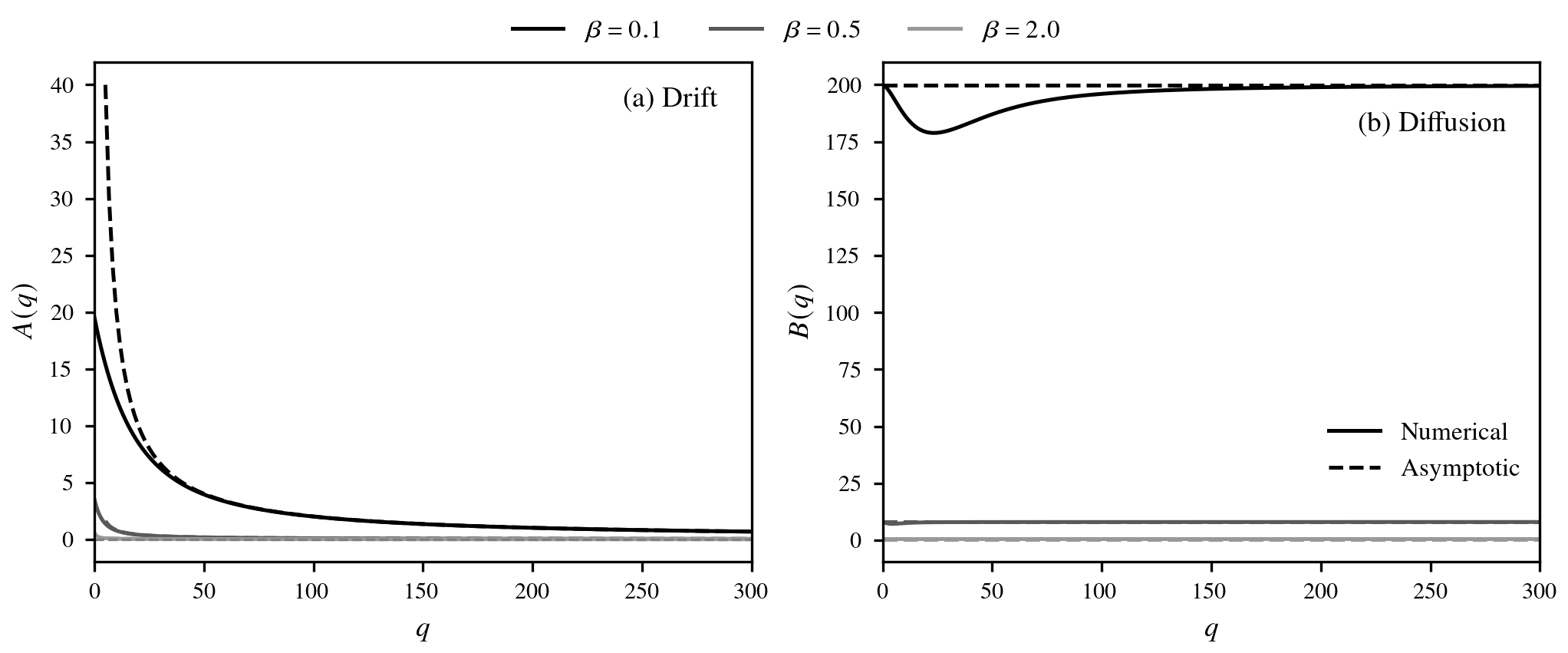}
	\caption{Drift and diffusion coefficients along with
		their asymptotic behavior.}
	\label{fig:driff_diff}
\end{figure*}

Fig. \ref{fig:driff_diff} shows the plot of these coefficients along with
their asymptotic behavior. The discrepancy at small $q$ originates
from the reflecting boundary at $q=0$, while excellent agreement
is recovered in the large-$q$ regime. The diffusion coefficient tends
to a constant solely determined by the parameter $\beta$. The drift
is much weaker, decaying as $1/q$. The outward drift originates
entirely from the degeneracy ratio $d_{r}/d_{q}=(2r+1)/(2q+1)$, which
slightly favors transitions toward larger irreducible representations.
In the asymptotic regime this bias is weak and scales as $1/q$. Far
from the boundary at $q=0$, the reflecting boundary, no longer influences
the dynamics at leading order. Numerical evaluation confirms that
the diffusion coefficient rapidly approaches the constant predicted
by the asymptotic analysis.

Since $A(q)=B/q$, the corresponding Itô stochastic equation is,

\begin{equation}
dq=\frac{B}{q}\,dn+\sqrt{B}\,dW_{n},
\end{equation}

which is equivalent, after a rescaling of time, to a one-dimensional
Bessel--type diffusion \cite{Karatzas_1998}. Standard properties of the Bessel process then
give the asymptotic evolution of the first two moments,

\begin{eqnarray}
	\langle q\rangle & \approx & \sqrt{q_{0}^{2}+2Bn}.\\
	\mathrm{Var}(q) & \approx & Bn
\end{eqnarray}

\begin{figure*}[t]
	\includegraphics[width=1\textwidth]{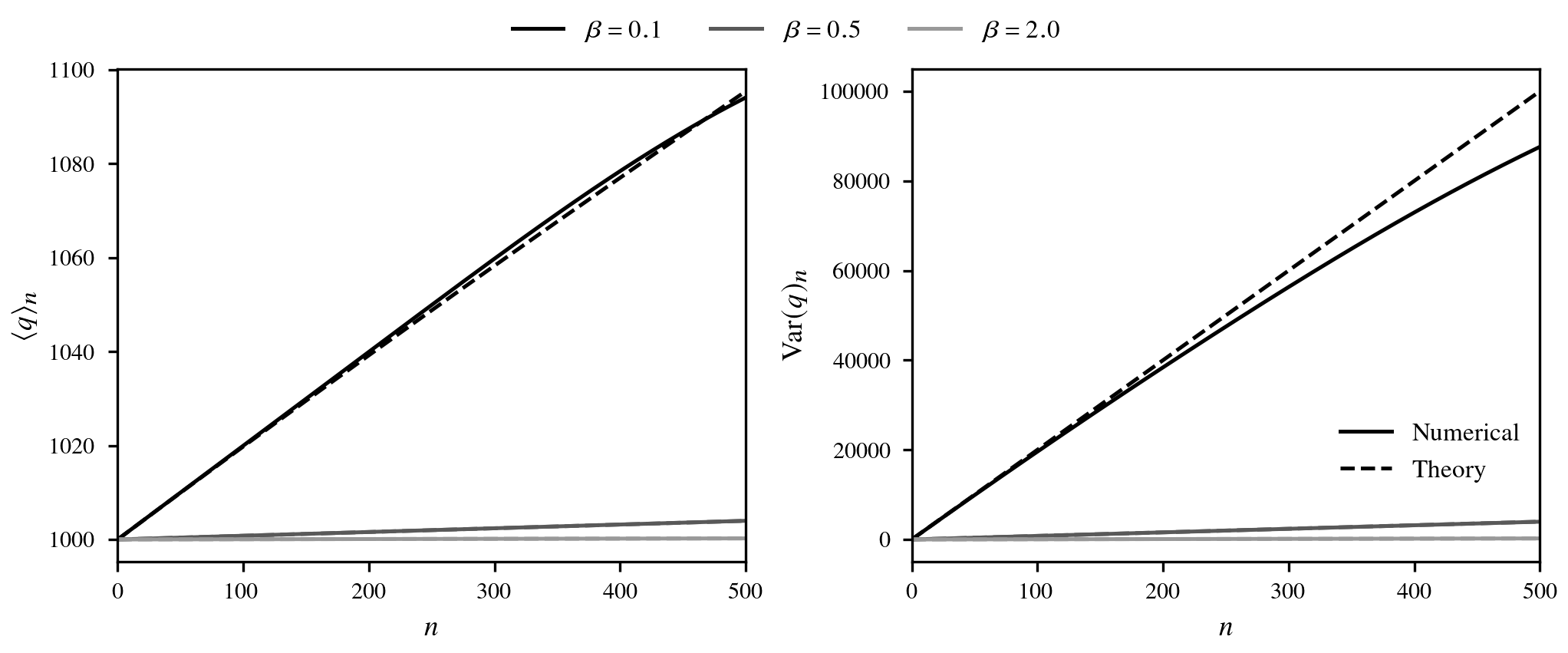}
	\caption{Numerical evolution of the first
		two moments with the asymptotic predictions}
	\label{fig:avg_r_var}
\end{figure*}

Fig. \ref{fig:avg_r_var} compares the numerical evolution of the first
two moments with the asymptotic predictions. The mean representation
follows the predicted square-root growth remarkably well over the
entire parameter range considered. The variance likewise exhibits
the expected linear growth for intermediate and large values of $\beta$.

For very small $\beta$, however, the numerical variance gradually
departs from the asymptotic prediction. In this regime the transition
kernel is broad and individual measurements induce comparatively large
jumps in representation space. Consequently, higher--order corrections
to the continuum (Kramers--Moyal) approximation and residual finite--lattice
effects become noticeable. In this regime the detector allows long--range transitions, so the
assumptions underlying the second--order Kramers--Moyal truncation
become less accurate.

Throughout the simulations the initial
state is chosen sufficiently far from the reflecting boundary at $q=0$
to minimize boundary effects. 

Thus, repeated covariant measurements induce an effective diffusion
on the representation graph whose transport coefficients are determined
entirely by the symmetry of the measurement and the detector bandwidth.
The continuum Fokker--Planck description therefore provides a coarse--grained
dynamical picture of measurement--induced transport in representation
space.

\section{Discussion and outlook}

In this work we developed a representation--theoretic description of covariant quantum measurements in which repeated measurements generate stochastic dynamics on the space of irreducible representations. The construction is constrained by the covariance of the measurement and the intertwining structure of the underlying symmetry group. For symmetry--invariant initial states, the averaged measurement channel closes on the irrep populations and becomes a classical Markov process whose transition probabilities are determined by the detector coefficients.

The general formalism was formulated for arbitrary compact Lie groups. Explicit calculations were carried out for SU(2), where the Clebsch--Gordan decomposition allows the transfer kernel to be evaluated analytically. By introducing an exponentially decaying detector spectrum consistent with the asymptotic behavior suggested by Sugiura's analyticity theorem, we obtained a closed--form transition kernel satisfying the normalization conditions required for a covariant POVM. The resulting Markov chain is reversible and admits an invariant measure; however, for the infinite representation graph this measure is not normalizable, indicating the absence of a stationary equilibrium distribution.

The continuum limit of the discrete walk leads naturally to a Fokker--Planck equation whose drift and diffusion coefficients are determined by the detector bandwidth and the representation--theoretic degeneracy factors. Numerical simulations show agreement with the continuum predictions in the regime where the Kramers--Moyal approximation is expected to hold.

The framework presented here suggests several natural extensions. Different detector spectra may produce qualitatively different transport processes, while finite representation spaces or confining detector profiles may generate normalizable stationary distributions. The same construction can also be applied to other compact symmetry groups, where the corresponding intertwining coefficients replace the Clebsch--Gordan coefficients used for SU(2). More generally, the present work suggests that covariant quantum measurements may be viewed as stochastic processes on representation graphs, providing a connection between quantum measurement theory, harmonic analysis on groups, and transport on discrete representation spaces.

\appendix

\section{Useful Clebsch--Gordan identities}

We use Condon--Shortley convention for the Clebsch--Gordon (CG) coefficients,
$C_{jm;JM}^{j^{\prime}m^{\prime}}\equiv\langle jm;JM|j^{\prime}m^{\prime}\rangle$
with the orthogonality relations,

\begin{eqnarray}
	\sum_{m,M}C_{jm;JM}^{j^{\prime}m^{\prime}}C_{jm;JM}^{k^{\prime}n^{\prime}} & = & \delta_{j^{\prime}k^{\prime}}\delta_{m^{\prime}n^{\prime}},\\
	\sum_{j^{\prime},m^{\prime}}C_{jm;JM}^{j^{\prime}m^{\prime}}C_{jn;JN}^{j^{\prime}m^{\prime}} & = & \delta_{mn}\delta_{MN}.
\end{eqnarray}

In terms of 3j--symbols,

\begin{eqnarray}\label{eq:3j_id}
	\sum_{m,M}\left(\begin{array}{ccc}
		j & J & j^{\prime}\\
		m & M & -m^{\prime}
	\end{array}\right)\left(\begin{array}{ccc}
		j & J & k^{\prime}\\
		m & M & -n^{\prime}
	\end{array}\right) & = & \frac{\delta_{j^{\prime}k^{\prime}}\delta_{m^{\prime}n^{\prime}}}{2j^{\prime}+1},\\
	\sum_{j^{\prime},m^{\prime}}(2j^{\prime}+1)\left(\begin{array}{ccc}
		j & J & j^{\prime}\\
		m & M & -m^{\prime}
	\end{array}\right)\left(\begin{array}{ccc}
		j & J & j^{\prime}\\
		n & N & -m^{\prime}
	\end{array}\right) & = & \delta_{mn}\delta_{MN}.
\end{eqnarray}

Also,
\begin{equation}
	C_{jm;JM}^{j'm'}=(-1)^{j-m}\sqrt{\frac{2j^{\prime}+1}{2J+1}}C_{jm;j'(-m')}^{J(-M)}\label{eq:CG_flip},
\end{equation}
and,

\begin{equation}\label{eq:CG_perm}
	\left(\begin{array}{ccc}
		j & J & j^{\prime}\\
		n & M & m^{\prime}
	\end{array}\right)=(-1)^{j+J+j'}\left(\begin{array}{ccc}
		j^{\prime} & J & j\\
		m^{\prime} & M & n
	\end{array}\right).
\end{equation}

\subsection{$\sum_{m^{\prime},M}C_{jn;JM}^{j^{\prime}m^{\prime}}C_{km;JM}^{j'm'}$}

Here the summation is over the detector index $M$ and the final magnetic quantum number $m^{\prime}$, which does not correspond to one of the standard Clebsch--Gordan orthogonality relations. Using 3j--symbols,

\begin{eqnarray}
	C_{jn;JM}^{j^{\prime}m^{\prime}} & = & (-1)^{j-J+m^{\prime}}\sqrt{2j^{\prime}+1}\left(\begin{array}{ccc}
		j & J & j^{\prime}\\
		n & M & -m^{\prime}
	\end{array}\right)\\
	C_{km;JM}^{j'm'} & = & (-1)^{k-J+m^{\prime}}\sqrt{2j^{\prime}+1}\left(\begin{array}{ccc}
		k & J & j^{\prime}\\
		m & M & -m^{\prime}
	\end{array}\right),
\end{eqnarray}

we first replace the dummy summation variable $m^{\prime}\rightarrow  -m^{\prime}$. Now using the symmetry of the Wigner 3j--symbols under the interchange of the first and third columns (\ref{eq:CG_perm}), and applying the standard orthogonality relation (\ref{eq:3j_id}), we obtain

\begin{equation}
	\sum_{m^{\prime},M}C_{jn;JM}^{j^{\prime}m^{\prime}}C_{km;JM}^{j'm'}=(2j^{\prime}+1)\frac{\delta_{jk}\delta_{mn}}{2j+1}\label{eq:CG_for_MM_1}.
\end{equation}

\subsection{$\sum_{m,m^{\prime}}C_{jm;KQ}^{j'm'}C_{jm;JM}^{j'm'}$}

Using Eq. \ref{eq:CG_flip}, we can easily get,
\begin{equation}
	\sum_{m,m^{\prime}}C_{jm;KQ}^{j'm'}C_{jm;JM}^{j'm'}=\frac{2j^{\prime}+1}{2J+1}\delta_{JK}\delta_{QM}\label{eq:diff_detector_ind}
\end{equation}

\section{Hilbert--Schmidt orthogonality of the intertwiners}\label{app:hs_ortho}

As, 
\begin{equation}
	T_{M;jj'}^{(J)}=\sum_{m,m'}C_{jm;JM}^{j'm'}|j'm'\rangle\langle jm|.
\end{equation}

we get,
\begin{equation}
	T_{Q;kk'}^{(K)}T_{M;jj'}^{(J)\dagger}=\delta_{jk}\delta_{j^{\prime}k^{\prime}}\sum_{m,m'}C_{km;KQ}^{k'm'}C_{jm;JM}^{j'm'}.
\end{equation}

Using Eq. \ref{eq:diff_detector_ind}
\begin{equation}
	\mathrm{Tr}\left(T_{Q;kk'}^{(K)}T_{M;jj'}^{(J)\dagger}\right)=\frac{d_{j^{\prime}}}{d_{J}}\delta_{jk}\delta_{j^{\prime}k^{\prime}}\delta_{JK}\delta_{QM},
\end{equation}

where $d_{\ell}=2\ell+1,\quad \ell=j^{\prime},J$, denotes the dimension of the SU(2) irrep of spin $\ell$.

\section{Completeness of the measurement operators}\label{app:M_complete}

Using the definition of measurement operator,
\begin{equation}
M_{g}=\sum_{J,j,j',M,N}c_{N;jj'}^{(J)}D_{MN}^{(J)}(g)T_{M;jj'}^{(J)},
\end{equation}

after Peter-Weyl orthogonality, we can get,

\begin{equation}
	{\bf I}=\int dgM_{g}^{\dagger}M_{g}=\sum_{J,j,j',M,N}\frac{1}{d_{J}}\sum_{k,k'}c_{N;jj'}^{(J)}\Big[c_{N;kk'}^{(J)}\Big]^{*}\Big[T_{M;jj'}^{(J)}\Big]^{\dagger}T_{M;kk'}^{(J)},
\end{equation}

with,

\begin{equation}
T_{M;kk'}^{(J)}=\sum_{m,m'}C_{km;JM}^{k'm'}|k'm'\rangle\langle km|.
\end{equation}

Finally using Eq. (\ref{eq:CG_for_MM_1}) we get,

\begin{equation}
\sum_{J,j,j',N}\frac{d_{j^{\prime}}}{d_{J}d_{j}}|c_{N;jj'}^{(J)}|^{2}\Pi_{j}=\bf{I}.
\end{equation}

\bibliographystyle{unsrt}
\bibliography{groupcov}

\end{document}